\documentclass[preprint,eqsecnum,superscriptaddress,nofootinbib,showpacs,amsmath,amssymb,tightenlines]{revtex4}

 \usepackage{graphicx}

\newcommand{\del}[0]{\partial}

\newcommand{\hateq}{\widehat{=}}
\newcommand{\D}{{\mathcal{D}}}
\newcommand{\Lie}[0]{{\cal L}}

 \makeatletter
 \newcommand{\pback}[1]{{
   \let\@rrow=\leftarrowfill
\mathchoice{\AIN@stemPullBack{#1}{\@rrow}}{\AIN@stemPullBack{#1}{\@rrow}}
     {\AIN@indxPullBack{#1}{\@rrow}}{\AIN@indxPullBack{#1}{\@rrow}}}
   \vphantom{#1}}

 \newcommand{\AIN@stemPullBack}[2]{
   \vtop{\mathsurround=0pt
   \ialign{##\crcr$\textstyle{#1}\strut$\crcr
     \noalign{\kern-0.4ex\nointerlineskip}{\tiny#2}\crcr}}}

 \newcommand{\AIN@indxPullBack}[2]{
   \vtop{\mathsurround=0pt
   \ialign{##\crcr\hfil$\scriptstyle{#1}$\hfil\crcr
     \noalign{\kern+0.4ex\nointerlineskip}{\tiny#2}\crcr}}}

\makeatother

\newcommand{\man}{{\mathcal{M}}}
\newcommand{\area}{a_\Delta}

\def\S{\mathcal{S}}

\def\beq{\begin{equation}}
\def\eeq{\end{equation}}
\def\bea{\begin{eqnarray}}
\def\eea{\end{eqnarray}}

\def\l{\ell}

\def\ba{\begin{eqnarray}}
\def\ea{\end{eqnarray}}
\def\be{\begin{equation}}
\def\ee{\end{equation}}
\def\={\hateq}
\def\R{R}
\def\Rt{\tilde\R}
\def\f{\frac}

\def\d{{\rm d}}
\preprint{\vbox{\baselineskip=12pt
 \rightline{gr-qc/0305044}
 \rightline{CGPG 2003-05/3}
\rightline{ICN-UNAM-03-05} }}

\begin{document}

\title{Non-minimally coupled scalar fields and isolated horizons}

\author{Abhay Ashtekar}\email{ashtekar@gravity.psu.edu}
\affiliation{Center for Gravitational Physics and Geometry \\
Physics Department, Penn State, University Park, PA 16802, USA}
\affiliation{Erwin Schr\"odinger Institute, Boltzmanngasse 9, 1090
Vienna, AUSTRIA}
\author{Alejandro Corichi}\email{corichi@nuclecu.unam.mx}
\affiliation{Instituto de Ciencias Nucleares\\
Universidad Nacional Aut\'onoma de M\'exico\\
A. Postal 70-543, M\'exico D.F. 04510, MEXICO}
\author{Daniel Sudarsky}
\email{sudarsky@gravity.psu.edu}
\affiliation{Center for Gravitational Physics and Geometry \\
Physics Department, Penn State, University Park, PA 16802, USA}
\affiliation{Instituto de Ciencias Nucleares\\
Universidad Nacional Aut\'onoma de M\'exico\\
A. Postal 70-543, M\'exico D.F. 04510, MEXICO}

\begin{abstract}

The isolated horizon framework is extended to include
non-minimally coupled scalar fields. As expected from the analysis
based on Killing horizons, entropy is no longer given just by (a
quarter of) the horizon area but also depends on the scalar field.
In a subsequent paper these results will serve as a point of
departure for a statistical mechanical derivation of entropy using
quantum geometry.

\end{abstract}
%
\pacs{04070Bw, 0420Fy}
\maketitle

\section{Introduction}
\label{s1}

The notion of a weakly isolated horizon
\cite{prl,abf,afk,abl2,abl1} extracts from Killing horizons the
minimal properties required to establish the zeroth and the first
laws of black hole mechanics. Thus, the notion captures the idea
that the horizon itself is in equilibrium, allowing for dynamical
processes and radiation in the exterior region. The resulting
isolated horizon framework has had several applications: i) For
fundamental physics, in addition to extending black hole mechanics
from stationary situations \cite{rw,mh}, it has led to a
phenomenological model of hairy black holes \cite{cs,acs} which
accounts for many of their key qualitative features; ii) In
computational relativity, it has provided tools to extract physics
from numerical simulations of gravitational collapse and black
hole mergers at late times \cite{prl,dkss,bk}; and, iii) For
quantum gravity, it provided a point of departure for a
statistical mechanical calculation of entropy, based on quantum
geometry \cite{abck}.

These applications feature gravity \emph{minimally} coupled to
matter such as scalar fields, Maxwell fields, dilatons, Yang-Mills
fields and Higgs fields \cite{afk,cs,ac,abl2}. A common element in
all these diverse situations is that the first law always takes
the form,
$$ \delta E = \frac{1}{8\pi G}\,\, \kappa \delta \area + {\rm work} $$
suggesting that a multiple of the surface gravity $\kappa$ should
be interpreted as the temperature and a multiple of the horizon
area $\area$ as entropy. This is striking because, irrespective of
the choice of matter fields \emph{and their couplings to each
other}, the entropy depends \emph{only} on a single geometrical
quantity, $\area$, and is independent of the values of matter
fields or their charges at the horizon. On the other hand, using
Killing horizons, Jacobson, Kang and Myers \cite{jkm} and Iyer and
Wald \cite{iw} have analyzed general classes of theories, showing
that the situation would be qualitatively different if the matter
is non-minimally coupled to gravity. Now, \emph{the expression of
entropy depends also on matter fields on the horizon}.

Specifically, consider a scalar field $\phi$ coupled to gravity
through the action
\be
 {\S}[ g_{ab}, \phi] =\int \d^4x \sqrt{- g}\left[
\frac{1}{16\pi G} f(\phi) R - \frac{1}{2}  g^{ab} \del_a \phi
\del_b \phi - V(\phi)\right]\, , \label{Action1} \ee
where $R$ is the scalar curvature of the metric $g_{ab}$ and $V$
is a potential for the scalar field. Then, the analysis of
\cite{jkm,iw} predicts that the entropy is given by
\be\label{S} S = \frac{1}{4G\hbar}\, \left(  \oint f(\phi)\, \d^2V
\right)\,. \ee
where the integral is taken on any 2-sphere cross-section of the
horizon. If the scalar field takes a constant value $\phi_0$ on
the horizon, the proportionality to area \emph{is} restored,
$S=[{f(\phi_0)} /{4\ell_{\rm P}^2}]\, \area$, but  even in this
case the constant now depends on $\phi_0$. Thus, non-minimal
coupling introduces a \emph{qualitative} difference.

It is natural to ask if the isolated horizon framework can
incorporate such situations. Apart from extending that framework,
the incorporation would also serve three more specific purposes.
First, we will have a richer class of examples. In particular the
analysis based on Killing horizons requires a globally defined
Killing field which admits a bifurcate horizon. While such
solutions admitting scalar hair are known to exist if the
cosmological constant $\Lambda$ is non-zero \cite{lambda},
analytic work \cite{NoScalarHair} and numerical evidence \cite{ps}
suggests that such solutions do not exist if $\Lambda$ vanishes.
The analysis \cite{jl} of the initial value problem based on
isolated horizons, on the other hand, can be used to show that
solutions admitting weakly isolated horizons would exist at least
locally, whence the isolated horizon analysis would not trivialize
in the $\Lambda =0$ case. The second point is more technical. In
\cite{jkm,iw}, a large class of theories is considered but under
the assumption that the action depends only on the metric, the
curvature and matter fields; first order actions which depend also
on the gravitational connection are not incorporated. The isolated
horizon analysis \cite{afk,abl2}, on the other hand, is based on
first order actions. Hence, incorporation of non-minimal couplings
in this framework would add to the robustness of the final results
of \cite{jkm,iw}. Finally, the analysis based on Killing horizons
does not provide an action principle or a Hamiltonian framework
which can be used for a non-perturbative quantization. The
isolated horizon framework does, thereby paving the way for a
fully statistical mechanical treatment based on quantum gravity.

The purpose of this paper is to extend the isolated horizon
framework to incorporate non-minimally coupled scalar fields of Eq
(\ref{Action1}). We will find that the entropy is indeed given by
(\ref{S}); the main result of \cite{jkm,iw} is robust. In a
subsequent paper we will show that this analysis provides the
point of departure for quantum theory and the fully quantum
mechanical calculation assigns the same entropy to the isolated
horizon.

Since the couplings of matter fields among themselves do not play
a direct role in our analysis, in the rest of the paper the term
`non-minimal coupling' will refer to the couplings of matter
fields to gravity.

\section{Non-minimally coupled fields in the first order formalism}
\label{s2}

We will first recall the second and first order actions of
interest and then specify the first order action that will be used
in the rest of the paper. For simplicity, in the first part we
will omit surface terms and restore them only at the end.

Let us then start with the action:
 \be
 {\S}[g_{ab},\phi]\, = \, \int_{\man} \d^4\!x \sqrt{- g}\left[
\frac{1}{16\pi G} f(\phi) R  - \frac{1}{2}  g^{ab} \del_a \phi
\del_b \phi - V(\phi)\right]
 \label{Action2}
 \ee
on a 4-dimensional manifold $\man$. In general relativity, one is
often interested in the case $f(\phi) = 1+8\pi G\xi \phi^2$ with
$\xi$ a constant.  Then $\phi$ satisfies a non-minimally coupled
equation $\Box \phi + \xi R \phi - {\partial
V(\phi)}/ {\partial \phi} = 0$.
However, our analysis is not tied to this case.

As indicated in section \ref{s1}, in the isolated horizon
framework it is simpler to work in the first order formalism. The
first order action for non-minimally coupled scalar fields was
discussed in \cite{rc,mmuv}. Let us begin by recalling the main
difference from the simpler, minimally coupled theory. In the case
of {minimal} couplings, the first order action is given just by
replacing the scalar curvature term in the second order action by
an appropriate contraction of the tetrads with the curvature of
the connection \cite{art}. In the present case, there is an
additional term because derivatives of the scalar field appear
when one integrates by parts to obtain the equations of motion of
the connection. More precisely, if we work with tetrads $e^a_I$
and Lorentz connections $A_{aI}{}^J$, the first order action is
given by:
\be
 {\S}[e,A, \phi] =
 \int_{\man} \d^4\!x\,\,  e\, \left[ \left(
\frac{1}{16\pi G}\, f(\phi)\,  e^a_I  e^b_J F( A)_{ab}^{IJ}\right)
 - \frac{1}{2}\,K(\phi) \del_a \phi \del_b \phi\,\,
 e^a_I  e^b_J \eta^{IJ} - V(\phi)\right] \, ,
 \label{Action3}
 \ee
where
\be \label{G} K(y) = [1 + (3/16\pi G) (f'(y))^2/f(y)]\, .\ee
To show that this action is equivalent to the second order action
(\ref{Action2}), it suffices to solve the equation of motion for
the connection $A$, substitute the solution in (\ref{Action3}) and
show that the result reduces to (\ref{Action2}).

Variation with respect to $A$ yields the equation of motion for
the connection:
 \be
 D_a\,\left(f(\phi)\,  e\,   e^a_{[I}  e^b_{J]}\right)=0.
 \label{EqA1}
\ee
Assuming that the \emph{rescaled} tetrad $\hat e^a_I= ( p\,
e^a_I)$, with $p= 1/\sqrt{f(\phi)}$, is well-defined and
non-degenerate, it follows that $A$ is the unique Lorentz
connection compatible with $\hat{e}^a_I$. Substituting this
solution in the expression of the curvature, we obtain:
 \be
  e^a_I  e^b_J F_{ab}^{IJ}(A) \,=\, \left(\frac{1}{p^2}\right)
  \hat e^a_I \hat e^b_J F_{ab}^{IJ}(A) \, = \, f(\phi)\hat{R}
  \label{Rhat}
 \ee
where $\hat R $ is the scalar curvature of the metric $\hat
g_{ab}=\hat e_a^I \hat e_b^J \eta_{IJ}$. Hence, in terms of the
metric $ g_{ab}=  e_a^I  e_b^J \eta_{IJ}$, the action
(\ref{Action3}) reduces to the second order form
\be
 {\S}[{g}, \phi] \, = \, \int_{\man} \d^4\!x \sqrt{- g}\left[
\left(\frac{1}{16\pi G}\right) f(\phi) {\hat R} - \frac{1}{2}
K(\phi)  g^{ab} \del_a \phi \del_b \phi  - V(\phi)\right]
 \label{Action4}
 \ee
Finally, using the standard relation between the scalar curvatures
of ${g}_{ab}$ and $\hat{g}_{ab}$, we recover, up to a surface
term, the second order action (\ref{Action1}) we began with.%
\footnote{Since in the isolated horizon framework the horizon is
treated as a physical boundary, the surface term has to
be examined carefully. It is proportional to $\oint\, \d S^a
\partial_a f$. We will see in the next section that on an isolated
horizon $\phi$ is `time independent'. It then follows that the
horizon contribution to the surface term vanishes whence the two
action principles are in fact equivalent.}
Thus, when $\hat{e}^a_I$ is smooth and non-degenerate, the first
order action (\ref{Action3}) is equivalent to the more familiar
one. However, we wish to emphasize that, for our purposes, it is
the first order action that is fundamental and this action as well
as the Hamiltonian framework developed in this paper continue to
be well-defined even when $f(\phi)$ vanishes or $e^a_I$ becomes
degenerate.

To make contact with literature on isolated horizons, it will be
convenient to rewrite the first order action (\ref{Action3}) in
terms of forms. Let us define the two form
$$\Sigma^{IJ}:=\f{1}{2} {\epsilon^{IJ}}_{KL}\,e^K\wedge e^L $$
where $e^I_a$ are the co-tetrads so that $e^a_I\,e_b^J
=\delta^a_b\delta^J_I$. The action now takes the form,
 \ba
{\S}[e,A,\phi]\, &=&\, \int_{\man} \left(\frac{1}{16\pi G}
\,f(\phi)\, \Sigma^{IJ}\wedge F_{IJ}+\frac{1}{2} \,K(\phi)\,
{}^\star\d \phi\wedge \d\phi - V(\phi)\,{}^4\epsilon\right)\,
\nonumber\\
&-&  \f{1}{16\pi G}\,\int_{\partial\man} A^{IJ}\wedge
\Sigma_{IJ}\, ,\label{actionforms}
 \ea
where ${}^\star$ denotes the Hodge-dual, ${}^4\epsilon$ is the
volume 4-form on $\man$ defined by the tetrad $e^a_I$ and where we
have explicitly included the surface term that is needed to make
the action differentiable. In the remainder of this paper we shall
use this form of the action.

We conclude with two remarks on the second order action.\\
1.  Consider again the second order action (\ref{Action2}). In the
sector of the theory in which $f(\phi)$ is nowhere zero, we can
pass to a conformally related metric $\bar{g}_{ab} = f(\phi)
g_{ab}$ and to a new field $\varphi =F(\phi)$  where $F$ is
defined by
\be \label{F} F(x) =\int^x \left[\frac{1}{f(y)} + \frac{3}{16\pi
G}\, \left(\frac{f'(y)}{f(y)}\right)^2\, \right]^{1/2}\, \d y \,
.\ee
Then the action (\ref{Action2}) can be rewritten as:
\be
{\S}[\bar{g}_{ab}, \varphi] =\int_{\man} \d^4x \sqrt{-\bar{g}}
\left[ \frac{1}{16\pi G} \bar{R} - \frac{1}{2}\bar{g}^{ab}
\del_a \varphi \del_b \varphi - v(\varphi)\right] \label{Action5}
\ee
where the potential for the scalar field $\varphi$ is given by
$v(\varphi)=(1/f^{2}(\phi(\varphi)))V(\phi(\varphi))$. Thus, on
this sector, the theory is equivalent to a minimally coupled
scalar field $\varphi$ on $(M, \bar{g})$. In the commonly used
terminology, (\ref{Action2}) expresses the theory in the Jordan
conformal frame while (\ref{Action5}) expresses it in the Einstein
frame.\\
2.  How stringent is the restriction that $f(\phi)$ be everywhere
positive? The equations of motion following from (\ref{Action2})
are:
\ba  0&=& g^{ab}\, \nabla_a \nabla_b \phi + \f{1}{16\pi G}\,
    R \,\f{\partial{f(\phi)}}{\partial \phi}
    + \frac{\partial V(\phi)}{\partial \phi}\nonumber\\
    \f{f(\phi)}{8\pi G}\left(R_{ab} - \frac{R}{2}\, g_{ab}\right)
    &=& \nabla_a \phi \nabla_b \phi - \left(\f{1}{2} \nabla^c \phi
    \nabla_c \phi - \nabla^c \nabla_c f (\phi) + V(\phi)\right) g_{ab}
    \nonumber\\
    &&+ \nabla_a \nabla_b f (\phi) \ea
Therefore, if $f(\phi)$ were to vanish in an open set, the Einstein
tensor is undetermined there. Consequently, it is likely that the
Cauchy problem would not be well-posed. Indeed, if the potential
admits a minimum at $\phi = k$, a constant, then on an open set on
which $f(\phi)$ vanishes, $\phi = k$, and $g_{ab}$ \emph{any} metric
with $R=0$ would satisfy the field equations.  By contrast, if
$f(\phi)$ is nowhere zero, as we just saw, the field equations are
equivalent to those of minimally coupled scalar field and the Cauchy
problem is then well-posed. Therefore, from the standard viewpoint
one adopts in general relativity, the requirement that $f(\phi)$
does not vanish on an open set is physically quite reasonable.\\

\section{Boundary conditions and action principle}
\label{s3}

Let us first adapt the basic definitions to accommodate
non-minimal couplings.

A \emph{non-expanding horizon} $\Delta$ is a null, 3-dimensional
sub-manifold of $(\man, g_{ab})$, topologically $S^2\times R$ such
that:\\
i) The expansion $\Theta_{(\ell)}$ of every null normal $\ell$ to
$\Delta$ vanishes; \\
ii) The scalar field $\phi$ satisfies $\Lie_{\ell}\, \phi \= 0$; and \\
iii) Equations of motion hold on $\Delta$.\\
Here and in what follows $\=$ denotes equality restricted to
points of $\Delta$. The previous papers on isolated horizons
assumed, in place of ii),  that the matter stress-energy satisfies
a very weak energy condition which, through the Raychaudhuri
equation, implied that matter fields are Lie dragged by $\ell^a$.
Non-minimally coupled scalar fields violate even that energy
condition. Therefore, we directly assume ii) which, it turns out,
suffices for our purposes. (This is the only change in the basic
definitions needed to accommodate non-minimal couplings.) This
condition captures the intuitive idea that, since the horizon
$\Delta$ is in equilibrium, the scalar field should be
time-independent on $\Delta$. The definition also implies that the
intrinsic (degenerate) metric $q_{ab}$ on $\Delta$ is time
independent; $\Lie_{\ell}\, q_{ab} \= 0$.

As in the minimally coupled case, the space-time covariant
derivative $\nabla$ induces a natural derivative operator $\D$ on
$\Delta$ such that $\D_a q_{bc} \= 0$ and $\D_a \ell^b \= \omega_a
\ell^b$ for some 1-form $\omega_a$ on $\Delta$. While $\D$ is
canonical, the 1-form $\omega_a$ depends on the choice of the
(future-directed) null normal; under rescaling $\ell^a \rightarrow
\tilde\ell^a = f\ell^a$ we have $\tilde\omega_a = \omega_a + \D_a
\ln f$.

A \emph{weakly isolated horizon} $(\Delta, [\ell])$ is a pair
consisting of a non-expanding horizon $\Delta$, equipped with an
equivalence class of null normals $\ell^a$ satisfying\\
\centerline{$\Lie_\ell\, \omega_a \= 0$,}\\
where $\ell \approx \ell^\prime$ if and only if $(\ell^\prime)^a =
c\ell^a$ for a positive constant $c$. As in \cite{afk,abl2}, to
establish the zeroth and the first law of black hole mechanics, we
will not need the stronger notion of isolated horizons.

As in \cite{afk,abl2}, an immediate consequence of these boundary
conditions is the zeroth law. Since $\Lie_\ell\, \omega_a \=0$, by
the Cartan identity, we have:
\be 2 \ell^a \D_{[a} \omega_{b]} + \D_b (\omega_a \ell^a)\, \=
0\ee
For reasons discussed in detail in \cite{abf}, $D_{[a}
\omega_{b]}$ is proportional to $\epsilon_{ab}$, the natural
2-volume element on $\Delta$ satisfying $\epsilon_{ab} \ell^b \=
0$. Hence, we conclude
\be \D_a \kappa_{\ell} : = \D_a (\omega_a \ell^a) \= 0\quad
\hbox{\rm i.e.} \quad \kappa_{\ell} \= {\rm const} . \ee

To formulate the action principle, we fix a manifold $\man$
bounded by two (would be) space-like, partial Cauchy surfaces
$M^\pm$, an internal boundary $\Delta$ which is topologically
$S^2\times R$ (the would be weakly isolated horizon). (See Figure
1.) We will choose  a fixed equivalence class of vector fields
$[\ell^a_0]$ along the `$R$ direction' of $\Delta$. It is
convenient also to fix a flat tetrad and a connection at infinity
and an \emph{internal} Minkowski metric $\eta_{IJ}$ and an
\emph{internal} null tetrad $\ell^I, n^I, m^I, \bar{m}^I$ on
$\Delta$. A \emph{history} will consist of an orthonormal tetrad
$e^a_I$, a Lorentz connection $A_a^{IJ}$ and a
scalar field $\phi$ on $\man$ such that:\\
i) $M^\pm$ are space-like, partial Cauchy surfaces on $(M, g_{ab}
:= \eta_{IJ} \, e_a^I e_b^J)$;\\
ii) $e^a_I \ell^I \in [\l^a_0]$ on $\Delta$;\\
iii) $(\Delta, [\ell])$ is a weakly isolated horizon; and,\\
iv) the fields satisfy the standard asymptotic flatness conditions
at spatial infinity. (For further details, see \cite{afk} or
\cite{abl2}.)

\begin{figure}
  \includegraphics{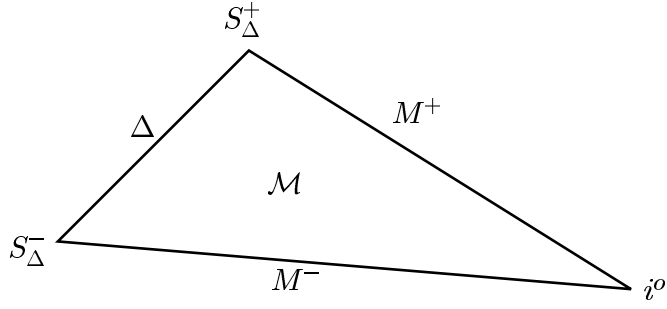}
  \caption{\label{f1}
  The region of space-time $\man$ under consideration has an
  internal boundary $\Delta$ and is bounded by two partial Cauchy
  surfaces $M^{\pm}$ which intersect $\Delta$ in the $2$-spheres
  $S^{\pm}_\Delta$ and extend to spatial infinity $i^{o}$. }
\end{figure}

In the variational principle, we fix the fields%
\footnote{Since we are working with a first order framework, we
will extend the action of the derivative operator $\nabla$ to
fields also with internal indices. Then $\nabla_a V_I = \partial_a
V_I + A_{aI}{}^J V_J$, where the flat derivative operator
$\partial$ on internal indices will be assumed to be compatible
with (i.e. annihilate) the internal tetrad on $\Delta$. Thus the
gauge freedom \emph{at} $\Delta$ is restricted.}
$e^a_I, A_a^{IJ}$ and $\phi$ on $M^\pm$. Since the boundary conditions
require ${\Lie}_\ell\, \phi \= 0$ and ${\Lie}_{\ell}\, \omega \=0$, in
each history that features in the variation, $\delta\phi \= 0$ and
$\delta \omega_a \= 0$. We will now use this information to show that
the first order action (\ref{actionforms}) leads to a well-defined
variational principle.

If we denote the variables $e_a^I, A^{IJ}, \phi$ collectively as
$\Psi$, we have:
 \be
 \delta {\S}[\Psi] = \int_{\cal M} \,E[\Psi]\,
 \delta \Psi + \int_{\del {\cal M}}  J [\Psi,\delta\Psi]
 \label{Var1}
 \ee
where $E[\Psi]= 0$ is the equation of motion for $\Psi$ and the
current 3-form $J$ is given by:
\be
 J [\Psi,\delta\Psi]= \frac{1}{16\pi G}\,\,\left( f(\phi)\,
  \Sigma^{IJ} \wedge  \delta  A_{IJ}\right)
 + K(\phi)\,{}^*\d \phi \wedge\delta \phi
 \label{u1}
 \ee
Thus, the action principle is well-defined if and only if the
integral of $J$ over the boundary of $\man$ vanishes. Now, since
all fields are kept fixed on $M^\pm$, $J$ itself vanishes there.
Similarly, the boundary conditions at spatial infinity ensure, as
usual, that the boundary term at infinity also vanishes.
Therefore, we only need to check the boundary integral on the
horizon $\Delta$. Since $\delta \phi \=0$, the second term in $J$
vanishes. To evaluate the first term, let us first calculate
$\delta A_b^{IJ}$ on $\Delta$. We first note that $A_a^{IJ}$ can
be expressed in terms of tetrads and $\omega_a$: Since
$$ \nabla_b \ell^a = \nabla_b(e^a_I \ell^I) = (\nabla_b e^a_I)
\ell^I - e^a_I A_b^{IJ} \ell_J\, , $$
using the equation of motion $\nabla_a \hat{e}^b_I = \nabla_a (p\,
e^b_I )\= 0$ which holds on $\Delta$ (because the definition of a
non-expanding horizon ensures that the field equations hold on
$\Delta$ ), we have
\be A_b^{IJ} \= 2 (\nabla_b \ln p) \ell^{[I} n^{J]} + 2 \omega_a
\ell^{[I} n^{J]} + C_b ^{IJ} + \ell_b U^{IJ} \ee
for some  $C_a^{IJ}$ and $U^{IJ}$ satisfying $C_a^{IJ} \ell_J \=
0$ and $U^{(IJ)} \= 0$. Now, since the internal tetrad is fixed
once and for all on $\Delta$, independently of the choice of the
history under consideration, variations of these internal vectors
vanishes. Similarly, since $\ell_b$ is the unique direction field
(in the cotangent space at any point of $\Delta$) which is
orthogonal to all tangent vectors to $\Delta$ in \emph{any}
history, $\delta \ell_b$ is necessarily proportional to $\ell_b$.
Therefore, the variation $\delta A_b^{IJ}$ is given by:
\be \delta A_b^{IJ} \= 2 (\nabla_b\, \delta \ln p\, +\delta\,
\omega_b)\, \ell^{[I} n^{J]} + \delta C_a^{IJ}\, +\, \delta U^{IJ}
\ell_b\label{deltaA} \ee

We can now substitute (\ref{deltaA})
in the surface term in $\delta S$. Using the fact that $\ell^a$ is a
null normal to $\Delta$ and the algebraic properties of $C_a^{IJ}$ and
$U^{IJ}$ on $\Delta$, we obtain:
\be \int_\Delta  J\, \=\, \frac{1}{8\pi G}\, \int_\Delta\, f(\phi)
\,  (\d( \delta \ln p) + \delta\omega)\wedge \epsilon = 0 \ee
where, in the last step, we have used $\delta \phi \= 0$ and
$\delta\omega \= 0$. Thus the surface term in (\ref{Var1})
vanishes because of the boundary conditions, whence the action
(\ref{actionforms}) is functionally differentiable. From
discussion of section \ref{s2} we know that the equations of
motion $E[\Psi] =0$ are the expected ones.

\section{Phase space and the first law}
\label{s4}

As in the previous papers on isolated horizons, we will use a
covariant phase space. Thus, our phase space $\Gamma$ will consist
of \emph{solutions} $(e_a^I, A_a^{IJ}, \phi)$ to the field
equations on $\man$ which are asymptotically flat and admit
$(\Delta, [\ell^a])$ as a weakly isolated horizon. (The precise
kinematical structure fixed on $\Delta$ is listed in Section
\ref{s3}.) As usual, the symplectic current is constructed from
the anti-symmetrized second variation of the action, where one can
now use the equations of motion \emph{but the variations are no
longer restricted to vanish on the initial and final surfaces
$M^\pm$}. To express the resulting symplectic structure in a
convenient form, it is convenient to introduce a scalar potential
$\psi$ of surface gravity $\kappa_{\ell}$ on $\Delta$ via:
\be \ell^a \nabla_a \psi = \kappa_\ell, \quad {\rm and} \quad
\psi\mid_{S^-_\Delta} = 0\ee
where, as in fig 1,  $S^-_\Delta$ is the 2-sphere on which
$\Delta$ intersects $M^-$. (Note that $\psi$ is independent of the
choice of vector $\ell^a$ within the equivalent class $[\ell]$.)
Then, the symplectic structure is given by:
\ba \label{sym}
 \Omega(\delta_1, \delta_2)\, =\!\! &&\f{1}{16\pi G} \int_M\,{\rm Tr}
 \left[\delta_1 (f\Sigma)\wedge \delta_2 A - \delta_2 (f\Sigma)\wedge
 \delta_1 A \right]\nonumber\\
 &+& \int_M\ K(\phi) \left[(\delta_1 \phi)\, \delta_2( {}^\star
 \d\phi) - (\delta_2 \phi)\, \delta_1({}^\star \d\phi)
 \right]\nonumber \\
 &+&\oint_{S_\Delta}\, \left[ \delta_1 (f\epsilon)\, \delta_2 \psi -
 \delta_1 (f\epsilon)\, \delta_2 \psi \right]\, ,
\ea
where $M$ is any partial Cauchy surface which intersects $\Delta$
in $S_\Delta$. Thus, as in the minimally coupled case
\cite{afk,abl2}, the symplectic structure has a surface term which
again comes from the `gravitational part' of the action --- the
first term in (\ref{actionforms})---  but now depends also on the
value of the scalar field on $\Delta$.

To obtain the first law, as in \cite{abl2} we need to make an
additional assumption: We will restrict ourselves to the case when
the weakly isolated horizons are of symmetry type II, i.e.,
\emph{axi-symmetric}. Thus, we will fix a rotational vector field
$\R^a$ on $\Delta$ (normalized so its closed orbits have affine
length equal to $2 \pi$) and restrict ourselves to the part of the
phase space where the fields $(q_{ab}, \omega_a, \phi)$ are all
Lie-dragged by $\R^a$. Note that the symmetry restriction is
imposed \textit{only} at $\Delta$; there is no assumption that the
fields in the bulk are axi-symmetric.

Following \cite{abl2}, let us first define the horizon angular
momentum. Consider any extension ${\Rt}^a$ of $\R^a$ on $\Delta$
to the bulk space-time $\man$ of which tends to an asymptotic
rotational Killing field at spatial infinity. The question is
whether motions along $\Rt^a$ induce canonical transformations on
the phase space and, if so, what its generating function $J_{\Rt}$
is. The simplest way to analyze this issue is to set $\delta_{\Rt}
= ({\Lie}_{\Rt}\, e,\, {\Lie}_{\Rt}\, A,\, {\Lie}_{\Rt}\,
\phi)$, and ask whether there exists a phase space function
$J^{\Rt}$ such that
\be \Omega (\delta,\, \delta_{\Rt}) = \delta J^{\Rt} \ee
for all tangent vectors $\delta$ to the phase space. The analysis
is completely analogous to that of \cite{abl2}. However, since one
has to use both the field equations obeyed by $(e, A, \phi)$ and
the linearized field equations satisfied by $(\delta e, \, \delta
A,\, \delta \phi)$, the calculation is significantly more
complicated because these equations are more involved now.
However, the final result is rather simple:
\be \Omega(\delta,\, \delta_{\Rt}) = \f{1}{16\pi G} \left[
 \oint_{S_\infty}\delta {\rm Tr} ({\Rt}^a A_a \,
 f\epsilon) \, - \, 2\oint_{S_\Delta} \, \delta (\R^a\omega_a\,
 f\epsilon) \right]\ee
where $\epsilon$ is the volume 2-form on the 2-sphere under
consideration. Hence, as is usual in generally covariant theories,
$J_{\Rt}$ consists only of surface terms. The term at infinity can
be shown to be the total angular momentum associated with $\Rt^a$.
It is then natural to interpret the surface integral at $\Delta$
as the isolated horizon angular momentum:
\be \label{J} J^{\R}_\Delta = -\f{1}{8\pi G} \oint_{S_\Delta} \,
\R^a\omega_a\, f\epsilon\, . \ee
Thus, the only difference from the minimally coupled case
\cite{abl2} is that volume 2-form $\epsilon$ there is now replaced
by $f\epsilon$.

We are now ready to obtain the first law. To introduce the notion
of energy, we have to consider vector fields $t^a$ in space-time
$\man$ representing time translations. Since we are dealing with a
generally covariant theory, what matters is only the boundary
values of $t^a$. It is clear that $t^a$ must approach an
asymptotic time translation at infinity and reduce to a symmetry
vector field representing time translations on the horizon:
\be t^a \= B_{(\ell, t)}\ell^a - \Omega_{(t)} \R^a \ee
where $B_{(\ell,t)}$ and $\Omega_{(t)}$ are constants on the horizon
and, as the notation suggests, $B$ depends not only on $t^a$ but also
on our choice of the null normal $\ell^a$ in $[\ell^a]$ (such that
$B_{(\ell, t)}\ell^a$ is unchanged under the rescalings of $\ell$ in
$[\ell]$.)  At infinity, all metrics tend to a universal flat metric
whence the asymptotic values of the symmetry vector fields are also
universal. At the horizon, on the other hand, we are in a strong field
region and the geometry is not universal. Therefore, there is
ambiguity in what one means by the `same' symmetry vector field in two
different space-times. For example, in the Kerr family, for the
`standard' time-translation, $\Omega_{(t)} \=0$ in the Schwarzschild
space-time but non-zero if the space-time has angular
momentum. Therefore, a priori we must allow for the possibility that
the horizon values of $t^a$ (i.e., the constants $B_{(\ell, t)},
\Omega_{(t)}$) may vary from one space-time to another. In the
numerical relativity terminology these are `live' vector fields. This
subtlety has nothing to do with non-minimal couplings and also arose
in all previous work on isolated horizons.

Again, the key question is whether there exists a phase space
function $E^{t}$ such that $\Omega(\delta, \, \delta_t) = \delta
E^{t}$. Using the equations of motion and their linearized
version, one can show that $\Omega(\delta, \, \delta_t) = \delta
E^{t}$ consists of two surface terms. Let us focus on the one at
the horizon. A long calculation yields:
\ba \Omega(\delta, \delta_t)\mid_{\Delta}\,  &=& \, \frac{1}{8\pi
G} \oint_{S_\Delta}\, \left[\kappa_{(t)}\delta(f\epsilon) \, -
\Omega_{(t)} \delta (\R^a\omega_a)\, f\epsilon \right]\nonumber\\
&=& \left[\f{\kappa_{(t)}}{8\pi G}\, \delta \oint_{S_\Delta} f\,
\epsilon\right] \, +\, \left[\Omega_{(t)} \delta J_\Delta^{\R}
\right] \ea
where $\kappa_{(t)}$ is the surface gravity (i.e. acceleration) of
the vector field $B_{(\ell, t)}\ell^a$ on $\Delta$. (The surface
term at infinity is precisely $\delta E_{\rm ADM}^{(t)}$.) Thus,
while the diffeomorphisms generated by $t^a$ give rise to a flow
on the covariant phase space, in general this flow may
\textit{not} be Hamiltonian. It is so if and only if there exists
a phase space function $E^{(t)}_\Delta$ such that
\be \label{1law}
 \delta E^{t}_\Delta = \left[ \f{\kappa_{(t)}}{8\pi G}\,
\delta \oint_{S_\Delta} f\, \epsilon\right] \, +\, \left[
\Omega_{(t)} \delta J_\Delta^{\R}\right] \, , \ee
i.e., if and only if the first law is satisfied. Thus, as in the
case of minimal coupling \cite{abl2}, the first law arises as a
necessary and sufficient condition for the flow generated by $t^a$
to be Hamiltonian.

As in \cite{abl2}, one can show that there exist infinitely many
vector fields $t^a$ for which the right side of (\ref{1law}) is an
exact variation and provide an explicit procedure to construct them.
Each of these vector fields generates a Hamiltonian flow and gives
rise to a first law. Thus, the overall structure is the same as in the
case of minimal coupling. However, there is a key difference in the
expression of the multiple of $\kappa_{(t)}/8\pi G$: while it is the
variation in the horizon area $a_\Delta$ in the case of minimal
coupling, now it is the variation of the integral of $f$ on a horizon
cross-section. Consequently, the entropy is now given by (\ref{S}).
As in \cite{abl2}, $E^{(t)}_\Delta$ and $\kappa_{(t)}$ depend on the
choice of the time translation $t^a$, \emph{while entropy does not}.

Finally, we note that if the horizon geometry fails to be
axi-symmetric, there is no natural notion of angular momentum.
However, we can still repeat the argument by seeking
`time-translation' vector fields $t^a$ with $t^a \= B_{(\ell,
t)}\ell^a$, the evolution along which is Hamiltonian. Such vector
fields exist. Thus, there is still a first law and the entropy
still given by (\ref{S}) and is again independent of the choice of
$t^a$.

We will conclude with some remarks.\\
i) It may first appear that the surface integrals in the
expressions (\ref{J}) of the horizon angular momentum and
(\ref{1law}) of the first law arise directly from the surface term
in the symplectic structure (\ref{sym}). This is not the case. In
fact in the detailed calculation the contributions of the surface
term in (\ref{sym}) to $\Omega(\delta,\, \delta_{\Rt})$ and
$\Omega(\delta,\, \delta_{t})$ vanish identically! The key surface
terms in (\ref{J}) and (\ref{1law}) come from integrations by part
of the bulk term in the symplectic structure, required to relate
the integrands to the field equations satisfied by $(e,A,\phi)$
and $(\delta e,\, \delta A,\, \delta\phi )$. The overall situation
is the same as in \cite{afk} for minimal coupling but these
calculations are now considerably more
complicated.\\
ii) Conceptually, it may appear strange that the flow generated by
$t^a$ is not always Hamiltonian. However, this is in fact the `rule'
rather than the `exception'. Consider the asymptotically flat
situation without boundaries and suppose we allow `live' vector
fields. Now, a live asymptotic time-translation vector field $t^a$ may
point in the same direction at infinity but may have a norm which
varies from space-time to space-time. (Indeed, it may even point in
different directions in different space-times.) The flow generated by
such vector fields in the phase space fails to be Hamiltonian in
general; it is Hamiltonian if and only of the asymptotic value of the
vector field is the same in all space-times; i.e., $t^a$ defines the
\emph{same} time translation of the fixed flat metric at infinity in
all space-times in the phase space. As mentioned above, in the case of
the isolated horizon, it is not a priori clear what the `same' time
translation on $\Delta$ means. The pleasant surprise is that it is the
first law that settles this issue.\\
iii) In the Einstein-Maxwell theory, one can exploit the black
hole uniqueness theorem to select a \emph{canonical} evolution
field $t^a$ on $\Delta$ of each space-time in the phase space
\cite{abl2}. There is then a canonical notion of horizon energy
---which can be taken to be the horizon mass--- and a canonical
first law. In the non-minimally coupled theory now under
consideration, it is no longer true that there is precisely a
2-parameter family of globally stationary solutions. Hence, one
can not select a canonical vector field or a canonical first law.
Consequently, although the expression of entropy is unambiguous,
that of horizon mass is not; we only have a $t$-dependent notion
of the horizon energy $E^{t}_\Delta$. Nonetheless, as with hairy
black holes \cite{acs}, one can extract physically useful
information from these expressions.\\
iv)  While we focused in this paper on 3+1 general relativity
without cosmological constant, it is straightforward to
incorporate 2+1 dimensional isolated horizons \cite{adw} and the
presence of cosmological constant \cite{abf}. In particular, our
results apply to the 2+1 dimensional black hole studied in
\cite{mz}, where, using Euclidean methods, entropy was found to be
$\frac{2}{3}\, (a_{\Delta}/4G\hbar)$. The `2/3' factor, which was
left as a puzzle, is naturally accounted for by the fact that
entropy depends also on the scalar field via (\ref{S}).

\section{Discussion}
\label{s5}

We have seen that, in presence of a weakly isolated horizon
internal boundary, there is a well-defined action principle and
Hamiltonian framework also for non-minimally coupled scalar
fields. While the overall structure is the same as in the
minimally coupled case, the first law is now modified, suggesting
that now the entropy is given by (\ref{S}). Thus we have extended
the main result of Jacobson, Kang and Myers \cite{jkm} and Iyer
and Wald \cite{iw} from stationary space-times to those admitting
only isolated horizons and  shown that the result holds also in
first order frameworks. This is however only a small extension
because whereas \cite{jkm} and \cite{iw} consider a very large
class of theories, here we considered only non-minimally coupled
scalar fields.

From the discussion of section \ref{s2}, it follows that we can
make a conformal transformation of the tetrad $e$ and a field
redefinition of $\phi$ to cast action (\ref{actionforms}) in to
that of the minimally coupled Einstein-scalar field theory with
tetrad $\bar{e}_a^I= \sqrt{f(\phi)} e_a^I$ and scalar field
$\varphi$ related algebraically to $\phi$ (via (\ref{F}).) It is
then natural to ask if our results would have been the same if we
had worked from the beginning in this minimally coupled `Einstein
frame' rather than the original non-minimally coupled `Jordan
frame'. One's first reaction may be that it is obvious that the
notions of angular momentum, energy and entropy of a field
configuration should not depend on whether we regard it as a
solution to the first theory or the second, whence the results
must be the same. However, this is not a priori clear. For
example, it is not self-evident that $(\Delta, [\ell])$ is even a
weakly isolated horizon for $(\man, \bar{g}_{ab})$. A more subtle
point is that there could be tension because even when a state is
shared by two theories, its properties \emph{can} depend on the
theory we choose to analyze it in. A striking example is provided
by the magnetically charged Reissner-Nordstrom space-times: While
they are stable in the Einstein-Maxwell theory, they are unstable
when regarded as solutions to the Einstein-Yang-Mills equations
\cite{rn}. In the isolated horizon framework, this difference can
be directly traced to the fact that the \emph{horizon} mass is
`theory dependent' \cite{acs}. More precisely, for a fixed
magnetically charged Reissner-Nordstrom space-time, the horizon
mass in the Einstein-Maxwell theory is the same as the ADM mass
while that in the Einstein-Yang-Mills theory is lower. The
difference can be radiated away, paving the way for an instability
in the Einstein-Yang-Mills theory \cite{acs}. The isolated horizon
framework even provides a qualitative relation between the horizon
area and the frequency of unstable modes in this theory.

It is therefore worthwhile to compare the results obtained here in
the Jordan frame with those one would obtain in the Einstein
frame. The main results can be summarized as follows:\\
\begin{itemize}

\item  $(\Delta, [\ell])$ is a non-expanding horizon in $(\man,
g_{ab})$ if and only if it is one in $(\man, \bar{g}_{ab} =
f(\phi) g_{ab})$. This follows because ${\Lie}_\ell\, \phi \= 0$.

\item The intrinsic horizon metrics are related by $\bar{q}_{ab} = f
q_{ab}$, and the volume 3-forms by $\bar{\epsilon} = f\,
\epsilon$. For any choice of the null normal $\ell^a$,
$\bar{\omega}_a = \omega_a +\f{1}{2} D_a \ln f$. Since the surface
gravity is given by $\kappa_\ell = \omega_a\ell^a$, and
${\Lie}_\ell\, \phi \= 0$ implies $\Lie_\ell\, f \= 0$, the two
surface gravities are equal: $\bar{\kappa}_\ell = {\kappa}_\ell$.

\item  In type II horizons, $\R^a\bar\omega_a \= \R^a\omega_a$
because ${\Lie_\R}\, \phi \= 0$. As a result, angular momenta
$J^{\R}_\Delta$ computed in the Einstein and Jordan frames agree.

\item  Diffeomorphisms generated by a vector field $t^a$ on $\man$
generate a Hamiltonian flow on the phase space in the Einstein
frame if and only if there exists a phase space function
$\bar{E}^{t}_\Delta$ such that

\be \label{1lawein}
 \delta \bar{E}^{t}_\Delta = \left[ \f{\kappa_{(t)}}{8\pi G}\,
 \delta \bar{a}_\Delta\right] \, +\, \left[ \Omega_{(t)}
 \delta J_\Delta^{\R}\right] \, ,
\ee
where $\bar{a}_\Delta$ is the horizon area in the Einstein frame.
Since the two volume 2-forms are related by $\bar\epsilon = f
\epsilon$, it follows that the right sides of (\ref{1law}) and
(\ref{1lawein}) are identical, whence the values of the entropy
and horizon energy calculated in the two frames are the same.

\end{itemize}

Thus, the main results are the same in the two frames; the situation
is different from that with the Einstein-Maxwell and
Einstein-Yang-Mills theories discussed above. This difference can be
traced back to the fact that while the phase spaces of those two
theories are quite different, \emph{in the case when $f$ is everywhere
positive}, the phase spaces of the minimally and non-minimally coupled
theories are naturally isomorphic.

Note however that the action (\ref{actionforms}), the boundary
conditions at $\Delta$, the symplectic structure (\ref{sym}), and
the vector field $\delta_t$ on the phase space are all
well-defined \emph{also in the case when $f$ vanishes on an open
set} of compact closure away from $\Delta$, so long as $K(\phi)$
remains everywhere
smooth.%
\footnote{If $f$ vanishes in an open set, the symplectic structure
at that point of the phase space acquires additional degenerate
directions, i.e., the notion of `gauge' is now enlarged. It would
be interesting to analyze whether the evolution is unique modulo
this extended gauge freedom even though, as pointed out in section
\ref{s2}, the Cauchy problem is ill-posed in the standard sense.}
Therefore, in the first order framework used here, the derivation of
the first law in the Jordan frame goes through even when one can not
pass to the Einstein frame through a conformal transformation.

At first sight, the appearance of the non-geometrical, scalar
field in the expression of entropy seems like a non-trivial
obstacle to the entropy calculation in loop quantum gravity
\cite{abck} because that approach is deeply rooted in quantum
geometry. In a subsequent paper we will use the Hamiltonian
framework developed here to pass to the quantum theory. One finds
that the non-minimal coupling does introduce conceptual changes in
the quantum geometry framework but one is again naturally led to a
coherent description of quantum geometry. The seamless matching
between the isolated horizon boundary conditions, the bulk quantum
geometry and the surface Chern-Simons theory which lies at the
heart of the calculation of \cite{abck} continues but the
statistical mechanical calculation now leads to the expression
(\ref{S}) of entropy.

\section*{Acknowledgments}
We would like to thank Bob Wald for suggesting that non-minimal
couplings offer an interesting test for the black hole entropy
calculations in loop quantum gravity, and Eanna Flanagan and Rob
Myers for discussions.  This work was supported in part by the NSF
grants PHY-0090091, CONACyT grant J32754-E and DGAPA-UNAM grant
112401, the Eberly research funds of Penn State, and the
Schr\"odinger Institute in Vienna.

\end{document}